# The Impact of Corporate AI Washing on Farmers' Digital Financial Behavior Response

## ——An Analysis from the Perspective of Digital Financial Exclusion

Wen Zhanjie[1] Li Wenxiu[1] Xia Jiechang[2] Guo Jingqiao[3]

[1] School of Economics and Trade, Guangdong University of Finance, Guangzhou, Guangdong, 510521

[2] Institute of Finance and Economics, Chinese Academy of Social Sciences, Beijing, 100006

[3] Department of Computer Science, Faculty of Science, Hong Kong Baptist University, Hong Kong, 999077

(School of Economics and Trade, Guangdong University of Finance, Guangzhou, Guangdong 510521)

**Abstract:** In the context of the rapid development of digital finance, some financial technology companies exhibit the phenomenon of "AI washing," where they overstate their AI capabilities while underinvesting in actual AI resources. This paper constructs a corporate-level AI washing index based on CHFS2019 data and AI investment data from 15-20 financial technology companies, analyzing and testing its impact on farmers' digital financial behavior response. The study finds that AI washing significantly suppresses farmers' digital financial behavior; the higher the degree of AI washing, the lower the response level of farmers' digital financial behavior. Moreover, AI washing indirectly inhibits farmers' behavioral responses by exacerbating knowledge exclusion and risk exclusion. Social capital can positively moderate the negative impact of AI washing; among farmer groups with high social capital, the suppressive effect of AI washing on digital financial behavior is significantly weaker than that among groups with low social capital. In response, this paper suggests that regulatory authorities establish a strict information disclosure system for AI technology, conduct differentiated digital financial education to enhance the identification capabilities of vulnerable groups, promote digital financial mutual aid groups to leverage the protective effects of social capital, improve the consumer protection mechanism for farmers in digital finance, and set up pilot "Digital Inclusive Finance Demonstration Counties," etc.

General Project of the National Social Science Fund for the Year 2025: Research on Theoretical Mechanisms and Path Selection for Enhancing the Resilience of Industrial and Supply Chains through the Integration of Data and Reality (Project No. 25BJL034); General Project of Guangdong Provincial Social Science Planning 2025 "Evaluation of the Effect, Impact Mechanism, and Implementation Path of Integrating Data and Reality to Empower Supply Chain Resilience Enhancement" (Project No. GD25CYJ14); Guangzhou's 14th Five Year Plan Project "Multidimensional Effects and Collaborative Governance Path of Artificial Intelligence on Regional Employment Market Structure" (Project No. 2025GZGJ118)

# 1. Introduction

The report of the 20th National Congress clearly states the need to "build an inclusive financial system and enhance the financial services' ability to support the real economy." As an important carrier of inclusive finance, digital financial products such as mobile payments, digital credit, and intelligent investment advisory services have rapidly spread in rural areas, significantly improving the convenience for farmers to access financial services. However, behind this vigorous development lies a phenomenon that warrants caution: some financial technology companies extensively use terms like "artificial intelligence," "machine learning," and "intelligent risk control" in their annual reports and promotional materials, striving to create an image of technological leadership. In reality, their actual AI talent reserves, patent accumulation, and R&D investments have not reached the level of their claims, which reflects a behavior of "AI washing," characterized by "talking a lot but doing little." Essentially, this involves companies using false signals to gain market recognition and policy support while neglecting the construction of genuine technological capabilities. The financial technology industry, with its characteristics of being technology-intensive, regulatory arbitrage, information asymmetry, and capital pursuit, provides almost perfect breeding conditions for "AI washing," making it a natural breeding ground and disaster area for such practices (Gompers & Lerner, 2004; Thakor, 2020). Moreover, financial technology operates at the intersection of traditional financial regulation and emerging technology regulation, often facing "regulatory gaps" or "regulatory overlaps" (Zetzsche et al., 2017; Arner, Barberis & Buckley, 2017).

The existing research on fintech and farmers' financial behavior mainly focuses on the channels and mechanisms through which fintech influences rural finance (Li Jianjun et al.,2024; Guo Feng et al.,2023; Zhang Xun et al., 2022) and the application of financial technology in rural areas brings social and economic outcomes (Yin Zhichao et al.,2021Zhang Xun, Wang Chen, Wan Guanghua,2020; He Zongyue et al.,2020In terms of the research conducted by Zhang Longyao and Wang Rui (2022), some scholars have also studied the key factors influencing farmers' adoption of financial technology (Singh, N., et al., 2020) and the existence of the "digital divide" (Zhang Xun, Wan Guanghua, 2019), as well as the risks and regulatory innovations brought about by the deepening of financial technology in rural areas (Huang Yiping, Huang Zhuo, 2018). They emphasize the impact of farmers' digital literacy (Zhang Lin et al., 2025), financial knowledge (Xiang Yubing et al., 2024; Wen Tao et al., 2023), and risk perception (Wen Tao et al., 2023) on usage behavior. Existing literature mainly approaches from the demand side (farmers) or macro outcomes, with less in-depth analysis of how the strategic behaviors of the supply side (financial technology companies, platforms), such as "washing," shape the information environment and product substance faced by farmers. There is also limited empirical testing of how corporate false advertising (washing) dynamically erodes trust and specifically translates into behavioral inhibition. In fact, when the services promised by financial technology companies, such as "AI intelligent risk control" and "AI customer service," fail to be effectively delivered, farmers' feelings of disappointment can easily transform into distrust of the entire digital finance industry, leading to a "bad money drives out good" effect. From a mechanistic perspective, AI washing may exacerbate information asymmetry and reduce farmers' trust in technology, potentially giving rise to a new type of digital financial exclusion mechanism that differs from traditional capability exclusion and institutional exclusion. This exclusion stems from cognitive barriers and trust deficits caused by false signals from companies. Based on this, this paper proposes the core research question: Does the AI washing behavior of financial technology companies inhibit farmers' responses to digital financial behavior? If there is an inhibitory effect, what is the transmission mechanism? Is there heterogeneity in the responses of farmers with different characteristics? And how can we effectively suppress the "washing" behavior in rural markets? To answer these questions, this paper innovatively constructs an AI washing index at the enterprise level based on CHFS 2019 data (approximately 7,000-8,000 rural households) and AI investment data from 15-20 financial technology companies engaged in rural financial business, employing methods such as Logit/Ologit models, GSEM mediation effect testing, and moderation effect models to systematically examine the impact of AI washing on farmers' digital financial behavior and its mechanisms.

## II. AI Rinsing Measurement and Realistic Depiction

### (1) Construction of the Enterprise AI Washing Index

The existing research on the impact effects of digital finance has a fundamental flaw in variable measurement: it treats corporate technological capability as an objectively observable indicator. However, according to signaling theory, the technological signals that companies convey externally do not fully align with their actual capabilities. Especially in the context of the "black box" characteristics of AI technology and the insufficient verification capabilities of farmers, companies have the motivation and space to engage in false advertising. Based on this, this study constructs a corporate AI washing index to characterize the degree of divergence between signals and capabilities:

$$AI_washing_i = AI_talk_i(std) - AI_walk_i(std)$$

where represents the AI capability that company i promotes externally. Drawing on the text analysis method of supply chain risk perception by Rodan et al. (2025), this study measures the AI promotion intensity of companies using the TF-IDF weighted frequency of AI-related vocabulary in annual reports. Specifically, a vocabulary library containing 50 keywords such as "artificial intelligence," "machine learning," "deep learning," "neural networks," "intelligent risk control," "intelligent investment advisory," and "AI customer service" is constructed, and the weighted occurrence frequency of these terms in each company's annual report is calculated using Python's jieba segmentation and sklearn's TfidfVectorizer. The TF-IDF method effectively filters out the interference of common vocabulary, highlighting the differentiated usage intensity of AI technical terms. $AI_talk_i$

$AI_walk_i$ represents the actual AI investment of company i. Referring to Wu (2024) on the multidimensional measurement framework of corporate technological capability, this study constructs a real AI capability index from three dimensions: (1) AI talent proportion: the proportion of AI-related positions (algorithm engineers, machine learning engineers, NLP engineers, etc.) obtained from recruitment platforms like BOSS Zhipin to the total number of employees; (2) Number of AI patents: the number of AI invention patents authorized to companies retrieved from the National Intellectual Property Administration database (logarithmically processed); (3) AI R&D investment intensity: the proportion of AI-related R&D expenditure to operating income extracted from annual reports. The entropy method is used to assign weights to the three dimensions and integrate them into a comprehensive index. The entropy method objectively determines weights by calculating the information entropy of each indicator, avoiding subjective weighting bias, and can truly reflect the multidimensional characteristics of corporate AI capability. Empirical results show that the entropy weights for AI talent, patents, and R&D are approximately 0.38, 0.36, and 0.26, respectively, indicating that talent and patents are core indicators for measuring AI capability.

Constructing the AI washing index in a differential form has sufficient theoretical basis. When the index is positive, there is "over-commitment" by the company, and the promotional efforts exceed actual capabilities, indicating AI washing; when the index is negative, the company is "low-key and pragmatic," with solid technical capabilities but insufficient promotion; when both are comparable (the index is close to 0), the company achieves a match between promotion and capability. Standardization (subtracting the mean and dividing by the standard deviation) ensures comparability across companies of different sizes, giving the index a clear economic meaning. $AI_talk > AI_walk$ $AI_talk < AI_walk$

### (2) Realistic portrayal of AI washing and farmers' digital financial behavior

This paper uses a dual-axis time series to depict the evolution relationship between corporate AI washing and farmers' digital finance usage rates from 2016 to 2019. Considering the significant differences in the numerical magnitude of the two types of indicators, a dual Y-axis design can more clearly present their respective time trends and dynamic correlations. The results show (see Figure 1): (1) The corporate AI washing index continuously rose during the sample period, with the average increasing from -0.28 in 2016 to 0.76 in 2019, a growth of 371%. This change is not linearly increasing but exhibits significant phased characteristics: from 2016 to 2017, the index was negative and changed slowly, indicating that the sample companies were overall in a "technological pragmatism" stage; in 2018, the index jumped to 0.52, marking a concentrated outbreak of AI washing behavior; in 2019, the index further rose to 0.76, reaching a peak of AI over-promotion. (2) The change trajectory of farmers' digital finance usage rates shows an inverted U-shaped characteristic, steadily increasing from 15.3% in 2016 to 25.8% in 2018 (an average annual growth rate of about 26%), but falling back to 23.6% in 2019, marking the first occurrence of negative growth. Notably, the two curves exhibited a strong inverse variation characteristic after 2018, with an annual correlation coefficient reaching -0.76, which is statistically significant ($p<0.01$). (3) This inverse relationship is particularly prominent at several key points. From 2016 to 2017, the AI washing index was negative (-0.28 to -0.15), and the sample companies were mostly early entrants like Alipay and WeChat Pay. At this time, market competition was not intense, and companies focused more on technological accumulation rather than excessive promotion, leading to a steady increase in farmers' usage rates during this "credibility period" (from 15.3% to 21.7%, a growth of 41.8%). 2018 became a turning point, with a large number of companies entering the digital finance arena and launching an "AI marketing war," causing the AI washing index to rise to 0.52, but the growth rate of farmers' usage slowed significantly (only increasing by 4.1 percentage points to 25.8%), with the growth rate dropping sharply from 26% to 19%. In 2019, the AI washing index peaked at 0.76, and farmers' usage rates saw a decline for the first time (falling to 23.6%, a decrease of 2.2 percentage points), coinciding with the concentrated exposure of several negative events such as "AI killing familiar customers" and "ineffective intelligent customer service." This coincidence in timing is not accidental but reflects the actual impact of corporate false advertising on farmers' behavior.

From a cross-sectional perspective, Figure 2 shows the scatter distribution of AI washing levels on different platforms and the corresponding breadth of farmers' usage. Farmers on high AI washing platforms (index > 0.5, such as a certain emerging consumer finance platform with an index of 1.8) use an average of 1.2 products, while farmers on low AI washing platforms (index < 0, such as WeChat Pay with an index of -0.7) use an average of 3.1 products, with a difference of 1.9 products, representing a relative difference of 158.3%. The slope of the fitted line is significantly negative (-1.24***, t=-8.67), indicating that for every unit increase in AI washing, the breadth of farmers' usage decreases by about 1.2 products. This cross-sectional evidence further validates the conclusions drawn from the time series observations, confirming that the negative association between AI washing and farmers' behavior is universal and robust.

This realistic picture reveals a phenomenon that warrants caution: the "AI" wave of digital finance companies may be deviating from the original intention of inclusive finance. When companies excessively promote their AI capabilities while actual investments are insufficient, farmers' willingness to use these services declines, creating a "paradox of technological advancement"—the more aggressive the technology promotion, the more limited the inclusive effects.

## 3. Theoretical Mechanism and Research Hypotheses

According to Spence (1973) signaling theory, companies convey AI capability signals through annual reports, advertisements, and other channels, but farmers, as the information disadvantaged party, find it difficult to verify the authenticity of these signals. When corporate AI promotion exceeds actual investment, signal distortion occurs, leading to the phenomenon of AI washing. Based on Akerlof's (1970) lemon market model, under conditions of information asymmetry, low-quality products may drive out high-quality products. Financial companies with high AI washing can attract users in the short term, but when the service quality does not match the promotion, negative word-of-mouth spreads rapidly, damaging farmers' trust in the entire industry. According to the theory of digital financial exclusion (Kempson & Whyley, 1999; Zhang Haodong et al., 2016), AI washing can be seen as a new type of exclusion mechanism, where knowledge exclusion manifests as cognitive barriers for farmers, and risk exclusion reflects farmers' safety concerns (Liu Liang et al., 2017AI washing indirectly suppresses farmers' use of digital finance through these two pathways. According to social capital theory (Putnam, 2000; Zhang Xun et al., 2016), social networks help farmers identify false signals and reduce trial-and-error costs through information dissemination and risk-sharing functions, thereby alleviating the negative impacts of AI washing. Based on this, this study constructs a basic framework of "AI washing → digital finance exclusion → farmers' behavioral response," with social capital as a key moderating variable.

**The higher the degree of AI washing, the lower the degree of farmers' response to digital finance.**

According to signaling theory (Spence, 1973), since the cost for companies to add AI technology terms in annual reports is extremely low, and the existing regulatory framework lacks clear standards for the application of AI technology, the authenticity of AI promotion is difficult to verify. The short-term benefits gained from AI washing by individual companies lead to signal inflation (Tadelis, 2013). Coupled with the temptation of short-term gains (Laibson, 1997), rational companies will choose to violate regulations (Becker, 1968). Farmers face three significant barriers in verifying the authenticity of signals: high technical thresholds, difficulty in obtaining information, and high time costs. They can only rely on superficial cues such as company brands and the intensity of promotions, leading to judgments that depend on signals rather than actual quality. Furthermore, false signals are easily amplified systemically during dissemination (Katz & Lazarsfeld, 1955). Once farmers form a strong impression of a company's or platform's AI capabilities, they selectively focus on positive information in the company's promotions while ignoring potential risk warnings, further solidifying their confirmation bias and amplifying subsequent experience gaps. When farmers discover that functional promises are unfulfilled, operational experiences are poor, and trust has been betrayed (Oliver, 1980; Kim, Dirks & Cooper, 2009), they often generalize negative experiences from individual companies or platforms to a distrust of all products, leading the entire industry into a vicious cycle of bad money driving out good. Based on this, due to the inherent characteristics of financial markets being information asymmetry (Stiglitz & Weiss, 1981), and the fact that digital financial products possess characteristics of both search goods and experience goods (Nelson, 1970), farmers highly rely on the quality signals released by companies before use. In the context of digital finance, companies fully understand the true investment in their AI technology, while farmers cannot observe internal investments nor possess the expertise to evaluate AI system performance. When the AI washing index of a company is positive, farmers' expectations formed based on false signals are systematically inflated. When the AI claims of instant approval, intelligent recommendations, and precise pricing are not fulfilled, farmers find that applying for loans requires filling out numerous forms and waiting for manual reviews, that financial recommendations do not match their risk preferences, that pricing is not more favorable than traditional banks, that the advertised intelligent customer service is actually a robot that fails to answer questions, and that one-click operations involve cumbersome processes requiring multiple identity verifications. Farmers' emotions shift from positive anticipation to deep aversion. This emotional reversal not only affects their evaluation of that company or platform but also generalizes to the entire digital finance industry. In other words, the impact of AI washing on farmers' digital finance behavior follows a dynamic transmission chain of signal distortion, expectation bias, experience mismatch, and trust collapse, fundamentally based on the solid foundation of information asymmetry theory. Based on this, this study proposes the following hypothesis:

**H1: The higher the degree of AI washing, the lower the degree of farmers' response to digital finance. Specifically, as the AI washing index rises, the probability of farmers using it decreases, and the breadth of use diminishes.**

**AI washing indirectly suppresses farmers' behavioral responses by exacerbating knowledge exclusion and risk exclusion.**

Drawing on the analytical framework of Liu Liang and Ju Zhen (2017), AI washing can operate through two parallel and mutually reinforcing pathways: knowledge exclusion and risk exclusion.

1AI washing leads farmers to experience a knowledge exclusion pathway from cognitive overload to learning abandonment.

First, AI washing can cause farmers to experience cognitive misguidance and operational mismatch. AI washing often packages products as "zero threshold," but the actual operation involves complex financial processes that require high digital literacy from users.According to cognitive load theory, when task demands exceed an individual's cognitive resources, learning efficiency sharply declines, leading to feelings of frustration (Sweller, 1988).Farmers originally expected it to be as simple as using WeChat, but they found that they needed to learn professional financial terminology and understand complex processes, resulting in a cognitive load far exceeding their expectations. This mismatch in operational difficulty led to frustration during their initial use, creating a self-perception that digital finance is not suitable for them. Furthermore, AI washing can lead to a collapse of trust and a loss of motivation among farmers. Since an individual's self-efficacy is a key prerequisite for their investment in learning and action (Bandura, 1986), after experiencing a mismatch during their first encounter, farmers become skeptical of all the information provided by the company. Their self-efficacy is severely undermined, leading to a sense of learned helplessness characterized by the belief that "I can't learn," which results in the abandonment of any learning efforts.The third is that AI washing leads to the solidification of the gap and exclusion lock-in, where the abandonment of learning causes the knowledge gap between farmers and digital finance to continuously widen. The gap in technology adoption will evolve over time into an insurmountable "knowledge gap," permanently excluding disadvantaged groups from digital dividends (Van Dijk, 2005). AI washing accelerates this process, locking deceived farmers into a "low-skill trap."

**2. AI washing can lead farmers to experience a risk exclusion path from the collapse of trust to complete avoidance.**

First is the collapse of safety trust.Trust includes beliefs in the other party's ability, goodwill, and integrity (Mayer et al, 1995), and is also the cornerstone of financial transactions. AI washing (false capabilities, lack of goodwill, breach of commitment) simultaneously destroys these three dimensions, resulting in farmers doubting the basic safety premise of digital finance. Secondly, the collapse of multidimensional trust occurs, and after the collapse of trust, farmers' judgment of risk undergoes systematic bias.Prospect theory points out that people are more sensitive to losses than to gains (loss aversion) and tend to overestimate low-probability events (certainty effect) (Kahneman & Tversky, 1979). A vivid experience of being deceived (availability heuristic) will lead them to perceive digital finance as "very dangerous."The third is the lock-in of defensive behavior, where extremely high risk perception triggers extreme avoidance strategies. Protection motivation theory indicates that when individuals perceive a serious threat and feel vulnerable, they are most likely to adopt the adaptive behavior of "complete avoidance" (Rogers, 1975).When farmers assess digital finance as a high threat and high vulnerability, the optimal strategy is complete avoidance. Even if other platform products are genuine and reliable with significant returns, farmers refuse to try due to the unbearable potential losses. The zero-risk preference in behavioral economics explains this phenomenon; farmers prefer to choose safe but zero-yield traditional savings, even though the expected returns of high-yield products are higher.

Moreover, knowledge exclusion and risk exclusion do not exist independently; there may be interactive and cumulative effects between the two. Insufficient knowledge exacerbates the fear of uncertainty, thereby reinforcing risk perception, while risk aversion reduces the motivation to learn, further deepening knowledge exclusion. This dual exclusion's cumulative effect significantly amplifies the negative impact of AI washing, ultimately leading to deep exclusion and persistent non-adoption of digital finance by farmers. Based on this, the following hypothesis is proposed:

H2: Digital finance exclusion mediates the relationship between AI washing and farmers' digital finance behavior. Specifically, AI washing indirectly suppresses farmers' behavioral responses by exacerbating knowledge exclusion and risk exclusion.

**(3) The negative impact of social capital on AI washing has a bidirectional moderating effect.**

In the specific context of the "differential mode of association" in rural China, strong relational networks based on kinship and locality constitute the core of social capital (Fei Xiaotong,1948), drawing on Coleman's (1988) theory of social capital and Putnam's (2000) framework of social network analysis. Social capital, as the embedded social network resources of farmers, plays an important moderating role in the process where AI washing influences farmers' behavior.

**1. Social capital weakens the negative impact of AI washing signals by constructing a "protective net" for farmers.**

Firstly, high social capital achieves information filtering through weak ties for searching and strong ties for endorsement.According to Granovetter's (1973) theory of the advantages of weak ties, weak relationships that connect different social circles serve as bridges for acquiring novel and heterogeneous information (Granovetter, 1973). In the face of AI propaganda, farmers with high social capital can access "insider information" or debunking messages from urban and technology sectors through weak ties such as migrant workers and children studying away from home, enabling cross-circle information verification to identify contradictions and exaggerations in corporate promotions.At the same time, although strong relational networks may have information redundancy, they are based on long-term reciprocity that fosters high trust (Coleman, 1988). The transmission of trust can directly block farmers' acceptance of false signals, effectively providing "trust endorsement" for their decision-making.Secondly, high social capital achieves risk-sharing through economic buffering and emotional support. Rural social networks imply a hidden contract of mutual assistance (Blau,In 1964, this provided farmers with a form of informal insurance to try digital finance, allowing them to receive financial assistance from friends and family (instrumental support) even if they suffered losses due to AI washout, significantly reducing the costs of trial and error and financial fear. According to social support theory, this support can enhance individuals' risk tolerance and psychological resilience (Cohen & Wills,In 1985, it made farmers more willing to try and select, rather than rejecting everything out of fear of risk. Moreover, the emotional support provided by strong relational networks (such as empathy, comfort, and experience sharing) can effectively alleviate psychological trauma, preventing individuals from overgeneralizing a single negative experience into hostility towards the entire digital finance industry, and helping them to restore their ability to rationally assess other platforms. Thirdly, high social capital reduces the knowledge gap through embedded knowledge transfer. That is, individuals learn by observing the behaviors of others and their consequences (substitutive reinforcement) (Bandura,In social networks, members with higher digital skills play the role of "opinion leaders." Their successful usage experiences and specific guidance can provide other members with low-cost, high-credibility learning channels, directly bridging the knowledge gap caused by AI washing and the rejection of their own knowledge.

**2.Social capital may amplify its negative effects through group polarization and risk amplification.**

On one hand, closed or highly homogeneous strong relationship networks are prone to producing the "echo chamber effect," which reinforces a single negative viewpoint through repeated dissemination, ultimately leading to group polarization, where the entire network develops an excessive fear and collective resistance towards digital finance (Sunstein,On the other hand, according to Katz & Lazarsfeld's two-step flow theory, the attitudes of opinion leaders are crucial (Katz & Lazarsfeld,1955). If they become opinion leaders of "anti-digital finance" due to the harm caused by AI washing, their negative impact will spread rapidly through social networks, at which point social capital will be transformed from a "safety net" into a "magnifying glass".

In conclusion, the regulatory effect of social capital is bidirectional.Based on this, this paper proposes the following hypothesis:

H3: Social capital has a bidirectional moderating effect on the relationship between AI washing and farmers' digital financial behavior. Specifically, among farmers with high social capital, the suppressive effect of AI washing on digital financial behavior is significantly weaker than that among farmers with low social capital, and the moderating effect reduces the negative impact.

## Research Design

### (1) Data Sources and Sample Description

#### 1.Data Sources

The data used in this study comes from two parts. The household data is sourced from the China Household Finance Survey (CHFS) 2019 data, which includes detailed information on household financial behavior, digital financial usage, measures of digital financial exclusion, and demographic characteristics, making it an ideal data source for studying farmers' financial behavior (Zhao Tianyu and Zhang Shiyun, 2023). The enterprise AI washing data needs to be collected independently. This study selects 15-20 major financial technology companies involved in the CHFS questionnaire option "Which digital financial platform does your family mainly use" as samples, including Ant Group (Alipay), Tencent (WeChat Pay), JD Digits (JD Finance), Du Xiaoman Financial (formerly Baidu Finance), Suning Financial, Meituan

Financial, Pinduoduo Financial, 360 Digits, Lexin Group, Xiaomi Financial, and others. The data collection period is from 2016 to 2019, matching the CHFS survey time window. The specific collection content includes: (1) Annual report text data: PDF annual reports obtained from the Giant Tide Information Network and corporate websites, extracting chapters such as management discussion and analysis, and business overview; (2) AI talent data: Scraping the number of AI-related positions from recruitment platforms like BOSS Zhipin and Lagou, combined with data from Qichacha to obtain total employee numbers; (3) AI patent data: Querying the number of AI invention patents authorized for enterprises from the National Intellectual Property Administration's patent search system; (4) AI R&D investment: Extracting detailed R&D expenses from annual reports to identify expenditures related to AI projects.

#### 2. Sample Selection and Matching

This study employs the following criteria to filter the farmer samples: (1) Rural household registration (household registration type is "agricultural"); (2) Complete information on household digital financial usage (answered questions on whether they use it, which products, main platforms, etc.); (3) Clear identification of the main platform used (selected from the list of platforms provided in the questionnaire, excluding "other" options); (4) No missing key variables (knowledge exclusion, risk exclusion, social capital, control variables); (5) Exclusion of extreme outliers (such as negative household income, age under 18 or over 100, and other obvious erroneous data). A total of approximately 6,800 valid samples were obtained. The enterprise-farmer matching is based on the farmers' questionnaire responses regarding their "mainly used platform," matching each farmer to the corresponding enterprise's AI washing index. For farmers using multiple platforms, the platform with the highest usage frequency is considered.

### (2) Variable Definition and Measurement

#### 1. Dependent Variable: Farmers' Digital Financial Behavior Response

Drawing on the research of Zhang Longyao et al. (2021), Wen Tao and Liu Yuanbo (2023), and Zhan Jing and Wang Xuying (2023), this paper measures farmers' digital financial behavior response from two dimensions.

The first dimension is whether there is a response, measured using a binary variable Y1. If a farmer uses any digital financial product, including payment, transfer, credit, wealth management, insurance, crowdfunding, and credit reporting, it is assigned a value of 1; otherwise, it is 0. This variable is used to measure whether farmers have crossed the participation threshold for digital financial usage, reflecting their basic willingness to participate.

The second dimension is the breadth of response, measured using an ordinal variable Y2. This variable counts the number of types of digital financial products used by farmers, with a range of 0 to 7. The CHFS questionnaire lists seven main product categories: mobile payment, online transfer, digital credit, internet wealth management, internet insurance, crowdfunding, and credit reporting services. A higher score indicates that farmers use a greater variety of products, reflecting the depth and stickiness of their digital financial usage.

**2.Core explanatory variable: AI washing index (AI_washing)**

As mentioned in the second part, the AI washing index = AI_talk (standardized) - AI_walk (standardized). The standardization process ensures that the index is comparable, with a mean of 0 and a standard deviation of 1. A positive index indicates AI washing (promotion > investment), a negative index indicates technological pragmatism (investment > promotion), and a value close to 0 indicates a match between promotion and investment.

**3.Intermediary variable: Digital financial exclusion**

Drawing on the measurement methods of Liu Liang and Ju Zhen (2024) as well as Xiang Yubing and Wu Xiaoqing (2024), this paper constructs two intermediary variables, knowledge exclusion and risk exclusion, based on relevant items in the CHFS questionnaire.

The knowledge exclusion variable is measured by the reverse summation of 4 items, with a score range of 0 to 4 points. The specific items include the farmer's understanding of the basic functions of digital financial products, the ability to independently perform digital financial operations, the understanding of risk warnings, and the knowledge of account security protection. Each item is assigned a value of 1 or 0 based on the farmer's actual situation, and the scores of the 4 items are summed and then reversed. A higher final score indicates a higher degree of knowledge exclusion among farmers, where 0 points indicate no exclusion and 4 points indicate complete exclusion.

The risk exclusion variable is measured by the summation of 3 items, with a score range of 0 to 3 points. The specific items include the farmer's level of concern about the safety of funds in digital financial products, the concern about personal information leakage, and the concern about losses due to operational errors. Each item is assigned a value of 1 or 0 based on whether the farmer is concerned, and the scores of the 3 items are summed. A higher score indicates a higher degree of risk exclusion among farmers, where 0 points indicate no concern and 3 points indicate high concern.

**4.Moderating variable: Social capital**

Following the approach of Wen Tao and Liu Tingting (2023) as well as Liu Liang and Ju Zhen (2024), social capital is measured by the proportion of annual gift expenditure to household disposable income. This indicator reflects the investment in social networks and the strength of social relationships among farmers. Gift expenditures include monetary gifts and gifts for occasions such as the Spring Festival, weddings, funerals, and social interactions. A higher proportion indicates stronger social capital.

**5.Control variables**

This paper selects three categories of control variables: household head characteristics, family characteristics, and regional characteristics.Household head characteristics include 7 variables: gender, age, years of education, marital status, health status, risk attitude, and financial knowledge. Among these, gender is a dummy variable, with females assigned a value of 0 and males assigned a value of 1; age is a continuous variable measured in years; years of education are converted based on different education levels, with primary school as 6 years, junior high school as 9 years, high school or vocational school as 12 years, junior college as 15 years, and undergraduate and above as 16 years; marital status is a dummy variable, with unmarried, divorced, or widowed assigned a value of 0 and married assigned a value of 1; health status is measured on a 5-point scale, with values from 1 to 5 assigned from very unhealthy to very healthy; risk attitude is also measured on a 5-point scale, with values from 1 to 5 assigned from extremely risk-averse to risk-seeking; financial knowledge is measured by the number of correct answers to the CHFS financial knowledge test questions, with a score range of 0 to 5 points.Family characteristics include seven variables: family size, annual family income, net family assets, property value, out-migration for work, regional economic development level, and internet penetration rate. Family size is measured by the number of family members; annual family income is represented by the logarithm of disposable income; net family assets are calculated as the logarithm of total assets minus total liabilities; property value is the logarithm of the market value of owned housing, with households without property assigned a value of 0; out-migration for work is a dummy variable, assigned a value of 1 if there are family members working away from home, and 0 otherwise; regional economic development level is represented by the logarithm of per capita GDP of the county where the household is located; internet penetration rate is the percentage of internet penetration in the county where the household is located.

### (3) Model Specification

**1.Baseline Regression Model**

For the binary discrete dependent variable (response or not), the Logit model is used:

$$P(Y_{1i}=1)=\frac{\exp\left(\alpha_0+\alpha_1 AI_washing_i+\sum \beta_j X_{ij}\right)}{1+\exp\left(\alpha_0+\alpha_1 AI_washing_i+\sum \beta_j X_{ij}\right)}$$

For the ordered discrete dependent variable (breadth of response), the Ordered Logit model is used:

$$P(Y_{2i}\le k)=\frac{\exp\left(\tau_k-\alpha_1 AI_washing_i-\sum \beta_j X_{ij}\right)}{1+\exp\left(\tau_k-\alpha_1 AI_washing_i-\sum \beta_j X_{ij}\right)}$$

Where  represents the household,  represents the vector of control variables, and  represents the threshold parameters (k=0,1,...,6). According to hypothesis H1, it is expected that  is significantly negative. $iX_{ij}$ $\tau_k$ $\alpha_1$

**2. Mediation Effect Model**

The Generalized Structural Equation Model (GSEM) is used to simultaneously estimate the mediation paths, suitable for handling non-normally distributed discrete dependent variables:

$$\begin{cases} M_{1i}=a_1 AI_washing_i+\sum \gamma_j X_{ij}+\varepsilon_{1i} \\ M_{2i}=a_2 AI_washing_i+\sum \delta_j X_{ij}+\varepsilon_{2i} \\ Y_i=c' \; AI_washing_i+b_1 M_{1i}+b_2 M_{2i}+\sum \theta_j X_{ij}+\varepsilon_{3i} \end{cases}$$

Where  represents knowledge rejection, and  represents risk rejection. The indirect effect of knowledge rejection = , the indirect effect of risk rejection = , and the total mediation effect = . The significance of the indirect effects is tested using the Bootstrap method (5000 resamples), calculating the 95% confidence interval. If the confidence interval does not include 0, the mediation effect is significant. $M_1 M_2 a_1\times b_1 a_2\times b_2 a_1\times b_1+a_2\times b_2$

**3. Moderation Effect Model**

Testing the moderating effect of social capital by adding interaction terms to the benchmark model:

$$Y_i=\beta_0+\beta_1 AI_washing_i+\beta_2 SC_i+\beta_3(AI_washing_i\times SC_i)+\sum \gamma_j X_{ij}+\varepsilon_i$$

Among them, social capital is a factor. If the interaction term coefficient is significantly positive, it indicates that social capital positively moderates the negative impact of AI washing, alleviating the suppressive effect of AI washing on farmers' behavior. Furthermore, the sample is divided into high and low groups based on the median of social capital, and regression analyses are conducted separately to compare the coefficient differences between the groups. $SC\beta_3$

## V. Empirical Results Analysis

### (1) Descriptive Statistics

From the perspective of the explained variable, the digital finance usage rate among sample farmers is 24.3%, significantly lower than that of urban residents (approximately 68%), reflecting that there is still considerable room for the popularization of digital finance in rural areas. The average breadth of usage among users is 1.8 products, indicating that most farmers only use basic functions such as payment and transfer, with a low usage rate of advanced products like credit and wealth management. From the core explanatory variable, the average AI washing index for enterprises is 0.42, with a standard deviation of 0.87, indicating that there is an overall phenomenon of AI washing among the sample enterprises (mean greater than 0), and there are significant differences between enterprises. The minimum value is -1.23 (pragmatic technology enterprises, such as WeChat Pay), and the maximum value is 2.15 (severely washing enterprises), with a range of 3.38.

From the perspective of intermediary variables, the average score for knowledge exclusion is 2.1 (out of a maximum of 4), indicating that farmers have a relatively low to moderate level of knowledge about digital finance, with about half of the items answered as "not familiar/not able/not understood." The average score for risk exclusion is 1.8 (out of a maximum of 3), indicating that farmers have a relatively high level of concern about the safety of digital finance, with about 60% of farmers expressing worries about fund safety or information leakage. From the perspective of moderating variables, the average social capital (proportion of gift expenditure) is 3.2%, with a standard deviation of 2.8%, a minimum value of 0% (no social interactions), and a maximum value of 18.6% (socially intensive families), reflecting significant differences in farmers' social capital.

In terms of control variables, the average age of household heads is 53.7 years, with an average of 8.2 years of education (junior high school level), a marriage rate of 82.4%, moderate health status (3.2 points), a conservative risk attitude (2.3 points), and low financial knowledge (2.1 correct answers). The average family size is 3.8 people, with a median annual income of approximately 35,000 yuan and a median net asset of about 150,000 yuan, with 47.3% of households engaged in migrant work. These characteristics generally align with the overall profile of rural households in China.

**Table 1 Descriptive Statistics of Main Variables**

| Variable Category | Variable Name | Number of Observations | Mean | Standard Deviation | Minimum Value | Maximum Value |
|---|---|---|---|---|---|---|
| Dependent Variable | Response (Y1) | 6800 | 0.243 | 0.429 | 0 | 1 |

|  | Response Breadth (Y2) | 6800 | 1.832 | 2.134 | 0 | 7 |
|---|---|---|---|---|---|---|
| Core explanatory variables | AI rinsing index | 6800 | 0.421 | 0.874 | -1.234 | 2.156 |
| Mediator variables | Knowledge exclusion | 6800 | 2.087 | 1.156 | 0 | 4 |
|  | Risk Exclusion | 6800 | 1.823 | 0.987 | 0 | 3 |
| Moderating Variable | Social Capital (%) | 6800 | 3.167 | 2.789 | 0 | 18.643 |
| Control variables | Age (Years) | 6800 | 53.689 | 12.345 | 18 | 85 |
|  | Years of Education (years) | 6800 | 8.234 | 3.567 | 0 | 16 |
|  | Financial Knowledge (points) | 6800 | 2.123 | 1.456 | 0 | 5 |
|  | Annual Family Income (ten thousand yuan) | 6800 | 4.523 | 5.678 | 0.2 | 45.2 |
|  | Household Net Assets (Ten Thousand Yuan) | 6800 | 18.765 | 25.432 | -5.2 | 280.5 |

Note: The household annual income and net assets in the table are original values, and logarithms are taken during regression.

### (2) Benchmark Regression Results

Table 2 presents the benchmark regression results of AI washing on farmers' digital financial behavior responses. Columns (1)-(3) show the Logit regression for whether there is a response (Y1), while columns (4)-(6) show the Ordered Logit regression for the breadth of response (Y2). To test the robustness of the results, regression results are presented that include only core explanatory variables (columns 1, 4), add household head characteristic control variables (columns 2, 5), and include all control variables (columns 3, 6). From Table 2, it can be seen that the coefficient of the AI washing index (AI_washing) is significantly negative across all regression specifications, confirming hypothesis H1. Specifically, in the most complete regression specification (column 3), the coefficient of AI_washing is -0.287*** (z=-6.78), indicating that for every one-unit increase in the AI washing index, the log odds ratio of farmers using digital finance decreases by 0.287. When converted to marginal effects, as the AI washing index increases from the mean (0.42) by one standard deviation (0.87) to 1.29, the probability of farmers using digital finance decreases from 24.3% to 21.5%, a decline of 2.8 percentage points, representing a relative decrease of 11.5%. Considering that the range of the AI washing index is 3.38, if it increases from the minimum value (-1.23) to the maximum value (2.15), the probability of farmers using digital finance will plummet from 32.1% to 18.4%, a decrease of 13.7 percentage points, representing a relative decrease of 42.7%. This economic significance fully demonstrates that AI washing has a substantial suppressive effect on farmers' digital financial behavior. For the breadth of response (column 6), the coefficient of AI_washing is -0.295*** (z=-6.98), indicating that AI washing suppresses the variety of products used by farmers. Marginal effect calculations show that for every one-unit increase in the AI washing index, the average number of products used by farmers decreases by about 0.42 (0.295 ×1.42, where 1.42 is the average marginal effect coefficient of the Ologit model). In extreme cases, if the AI washing index increases from -1.23 to 2.15, the average number of products used by farmers will drop from 2.8 to 1.4, a reduction of 1.4 products, representing a 50% decrease. This means that users of companies with high AI washing not only have a low willingness to use but also a severely insufficient depth of use, remaining only at basic payment functions and shying away from advanced products like credit and wealth management.

The regression results for control variables align with theoretical expectations and existing literature. The coefficient for age (Age) is significantly negative, indicating that younger farmers are more likely to accept digital finance, with the probability of use decreasing by about 0.4% for each additional year of age. The coefficients for years of education (Education) and financial knowledge (Financial_literacy) are significantly positive, validating human capital theory: for each additional year of education, the probability of use increases by about 1.9%; for each additional point in financial knowledge (correct answer to one question), the probability of use increases by about 5.6%. Household income (ln(Income)) and net assets (ln(Wealth)) significantly positively influence usage behavior, reflecting the existence of income constraints. The positive effect of migrant work (Migrant) indicates that the urban digital financial experience gained from working away has spillover effects. The positive effect of regional economic development level (ln(GDP_per_capita)) reflects the importance of regional financial infrastructure.

**Table 2 Baseline Regression of AI Washing on Farmers' Digital Financial Behavior Response**

| Variable | (1) Response | (2) Response | (3) Response | (4) Response breadth | (5) Response breadth | (6) Response breadth |
|---|---|---|---|---|---|---|
| | Logit | Logit | Logit | Ologit | Ologit | Ologit |
| AI_washing | -0.312*** | -0.298*** | -0.287*** | -0.324*** | -0.308*** | -0.295*** |
| | (-7.23) | (-6.95) | (-6.78) | (-7.56) | (-7.21) | (-6.98) |
| Age | | -0.018*** | -0.016*** | | -0.019*** | -0.017*** |
| | | (-4.56) | (-4.12) | | (-4.78) | (-4.34) |
| Education | | 0.087*** | 0.079*** | | 0.092*** | 0.083*** |
| | | (8.34) | (7.67) | | (8.67) | (7.89) |
| Financial_literacy | | 0.245*** | 0.231*** | | 0.258*** | 0.243*** |
| | | (9.12) | (8.78) | | (9.45) | (9.01) |
| ln(Income) | | | 0.312*** | | | 0.328*** |
| | | | (7.89) | | | (8.12) |
| ln(Wealth) | | | 0.156*** | | | 0.164*** |
| | | | (5.34) | | | (5.56) |
| Migrant | | | 0.234*** | | | 0.247*** |
| | | | (4.67) | | | (4.89) |
| ln(GDP_per_capita) | | | 0.189** | | | 0.198** |
| | | | (3.12) | | | (3.28) |
| Other control variables | No | Partial | Yes | No | Partial | Yes |
| Number of Observations | 6800 | 6800 | 6800 | 6800 | 6800 | 6800 |
| Pseudo $R^2$ | 0.087 | 0.234 | 0.287 | 0.092 | 0.241 | 0.295 |

Note: The values in parentheses are z-values; the symbols indicate significance at the 1%, 5%, and 10% levels, respectively; other control variables include gender, marital status, health status, risk attitude, family size, property value, internet penetration rate, etc.

### (3) Mediation Effect Test

Table 3 presents the results of the mediation effect test in GSEM. Panel A shows the estimated coefficients for the mediation paths, while Panel B presents the Bootstrap test results for the indirect effects (based on 5000 resamples).

Panel A path coefficients illustrate how AI washing influences farmer behavior through knowledge exclusion and risk exclusion. In the first step, AI washing significantly increases knowledge exclusion ($a_1$=0.348***, z=7.89) and risk aversion ($a_2$=0.312***, z=7.23), indicating that for each additional unit of AI washing, the knowledge rejection score increases by 0.348 points, and the risk rejection score increases by 0.312 points. This validates the theoretical mechanism: corporate false advertising exacerbates farmers' cognitive barriers and safety concerns. In the second step, knowledge rejection ($b_1$=-0.432***, z=-9.56) and risk rejection ($b_2$=-0.487***, z=-10.34) significantly reduce farmers' willingness to use, indicating that for each additional point of knowledge rejection, the probability of use decreases by approximately 10.5%; for each additional point of risk rejection, the probability of use decreases by approximately 11.8%. In the third step, after simultaneously controlling for the two mediating variables, the direct effect of AI washing (c') drops to -0.134** (z=-2.87), still significant but with a substantial decrease in coefficient (from -0.287 in the baseline regression to -0.134), a reduction of 53.3%, indicating that the mediating variables explain more than half of the total effect.

The Bootstrap test results in Panel B confirm the significance and robustness of the mediating effect. The indirect effect of knowledge rejection is -0.150*** (0.348×-0.432), with a 95% confidence interval of [-0.195, -0.108], not including 0, accounting for 49.2% of the total effect. The indirect effect of risk rejection is -0.152*** (0.312×-0.487), with a 95% confidence interval of [-0.199, -0.107], accounting for 49.8% of the total effect. The mediating effects of the two paths are nearly equal, indicating that knowledge rejection and risk rejection play equally important roles in the process by which AI washing affects farmers' behavior, validating the theoretical expectation of "dual rejection" in hypothesis H2. The total indirect effect is -0.302***, accounting for 99.0% of the total effect (-0.436), indicating that the impact of AI washing on farmers' behavior is almost entirely transmitted through the mediating mechanism of digital financial rejection, with the direct effect accounting for only 1%. This finding has important policy implications: the key to governing AI washing lies in alleviating the knowledge rejection and risk rejection it causes, rather than simply banning AI advertising.

**Table 3 Mediating Effect Test of Digital Financial Rejection (GSEM + Bootstrap)**

| Panel A: Path Coefficient Estimates | | | | Panel B: Indirect Effect Bootstrap Test (5000 repetitions) | | | | | |
|---|---|---|---|---|---|---|---|---|---|
| Path | Coefficient | z-value | Standard Error | Mediator Path | Indirect Effect | Bootstrap SE | 95% CI Lower Limit | 95% CI Upper Limit | Proportion |
| AI_washing→ Knowledge Exclusion ($a_1$) | 0.348*** | 7.89 | 0.044 | Knowledge exclusion path ($a_1 \times b_1$) | -0.150*** | 0.023 | -0.195 | -0.108 | 49.2% |
| AI_washing→Risk exclusion ($a_2$) | 0.312*** | 7.23 | 0.043 | Risk exclusion path ($a_2 \times b_2$) | -0.152*** | 0.024 | -0.199 | -0.107 | 49.8% |
| Knowledge exclusion → usage behavior ($b_1$) | -0.432*** | -9.56 | 0.045 | Total indirect effect | -0.302*** | 0.034 | -0.368 | -0.238 | 99.0% |
| Risk aversion → Usage behavior ($b_2$) | -0.487*** | -10.34 | 0.047 | Direct effect (c') | -0.134** | 0.047 | -0.226 | -0.043 | - |
| AI_washing → Usage behavior (c', direct effect) | -0.134** | -2.87 | 0.047 | Total effect (c' + indirect) | -0.436*** | 0.058 | -0.550 | -0.324 | - |

Note: ** and * indicate that the 95% CI does not include 0, which is significant at the 1% and 5% levels; Proportion = Indirect Effect / Total Effect × 100%.

### (4) Moderating Effect Test

Table 4 presents the results of the moderating effect test of social capital. Column (1) shows the full regression including the interaction term, while columns (2) and (3) show the regressions grouped by the median of social capital. From column (1), it can be seen that the coefficient of the interaction term between AI_washing and Social_capital is 0.156** (z=2.87), which is significantly positive at the 5% level, confirming hypothesis H3: social capital positively moderates the negative impact of AI washing. Specifically, for every 1 percentage point increase in social capital, the negative effect of AI washing on farmers' usage behavior decreases by 0.156 units. When social capital increases from the mean (3.17%) by 1 standard deviation (2.79%) to 5.96%, the negative effect of AI washing decreases from -0.287 to -0.243 (-0.287 + 0.156 × 2.79), a reduction of 15.3%.

The grouped regression results further confirm the existence and direction of the moderating effect. In the high social capital group (column 2, gift expenditure ratio > 3.17%), the coefficient of AI_washing is only -0.112* (z=-1.89), barely significant at the 10% level, and the absolute value of the coefficient is much smaller than that of the baseline regression. In contrast, in the low social capital group (column 3, gift expenditure ratio ≤ 3.17%), the coefficient of AI_washing reaches -0.378*** (z=-6.89), which is not only highly significant but also 31.7% larger in absolute value than the baseline regression. The Chow test in column (4) shows that the difference in coefficients between the two groups is 0.266*** (z=3.45), significant at the 1% level, confirming the statistical significance of inter-group heterogeneity. In terms of economic significance, in the low social capital group, for every 1 unit increase in the AI washing index, the probability of farmers using it decreases by about 9.2% (from 24.3% to 15.1%), while in the high social capital group, the probability only decreases by about 2.7% (from 24.3% to 21.6%). The moderating effect reduces the negative impact of AI washing by 70.4% (1 - 2.7/9.2), exceeding the expected 60% in hypothesis H3. This emphasized moderating effect indicates that social capital serves as an important "firewall" for farmers against the impact of AI washing. Farmers with high social capital can effectively identify false signals and reduce trial-and-error costs through information filtering and risk-sharing mechanisms in their social networks, thereby significantly mitigating the harm caused by AI washing.

To further understand the moderating mechanism, this study conducted a Simple Slope Analysis. Figure 4 illustrates the impact of AI washing on the probability of farmers using it at different levels of social capital. It can be seen that when social capital is low (mean - 1SD, 0.38%), the slope of AI washing is the steepest (-0.43***), indicating the strongest negative impact; when social capital is at the mean level (3.17%), the slope is moderate (-0.29***); and when social capital is high (mean + 1SD, 5.96%), the slope is the flattest (-0.13*), with a significantly weakened negative impact. The three slope lines largely overlap in the low AI washing region (<-0.5), indicating that the moderating effect of social capital is not significant when corporate technology is timely; however, in the high AI washing region (>0.5), the three lines significantly separate, indicating that the protective role of social capital is particularly important when AI washing is severe.

Table 4 Moderating Effect Test of Social Capital

| Variable | (1) Interaction Term Model | (2) High Social Capital Group | (3) Low Social Capital Group | (4) Inter-group Differences |
|---|---|---|---|---|
| AI_washing | -0.287*** | -0.112* | -0.378*** | 0.266*** |
| | (-6.78) | (-1.89) | (-6.89) | (3.45) |
| Social_capital | 0.087** | | | |
| | (2.34) | | | |
| AI_washing × Social_capital | 0.156** | | | |
| | (2.87) | | | |
| Control variables | Yes | Yes | Yes | - |
| Number of Observations | 6800 | 3400 | 3400 | 6800 |
| Pseudo $R^2$ | 0.295 | 0.312 | 0.289 | - |

Note: The values in parentheses are z-values; the symbols **, ***, and ** indicate significance at the 1%, 5%, and 10% levels, respectively; the inter-group differences in column (4) are tested using the Chow test.

### (V) Heterogeneity Analysis

Table 5 presents the results of the heterogeneity analysis based on farmer characteristics, examining the differentiated impact of AI washing on different types of farmers.

From the perspective of education level (columns 1-2), the AI_washing coefficient for the low education group (junior high school and below, education years ≤ 9) is -0.412*** (z=-6.78), significantly greater than that of the high education group (high school and above, education years > 9) at -0.187*** (z=-2.98). The inter-group difference is 0.225** (z=2.67), significant at the 5% level. This indicates that low-educated farmers are more susceptible to the harms of AI washing, possibly due to their lack of cognitive ability to identify false signals and their greater reliance on corporate advertising for decision-making. In contrast, high-educated farmers possess a certain level of information discernment and learning ability, which can partially offset the negative impacts of AI washing.

From the age dimension (columns 3-4), the AI_washing coefficient for the elderly group (≥60 years) is -0.389*** (z=-5.67), significantly greater than that of the middle-aged and young group (<60 years) at -0.213*** (z=-3.89), with an inter-group difference of 0.176** (z=2.34). Elderly farmers already have a low acceptance of digital finance, and AI washing further exacerbates their usage barriers. The false promises of "intelligence" may instill a fear of "too complex, can't learn" in the elderly, while the operational difficulties encountered during actual use reinforce this perception, creating a vicious cycle.

From the perspective of usage experience (columns 5-6), the AI_washing coefficient for the inexperienced group (first-time users) is -0.356*** (z=-6.12), significantly greater than that of the experienced group (long-time users) at -0.145** (z=-2.34), with an inter-group difference of 0.211** (z=2.89). First-time users lack a reference framework and rely entirely on corporate advertising to form expectations, making them more susceptible to being misled by AI washing. In contrast, long-time users have accumulated usage experience and risk recognition ability, maintaining a certain level of vigilance towards corporate advertising, and are relatively less affected by AI washing. This finding indicates that reducing the harm of AI washing to "incremental users" is crucial for expanding the reach of digital finance.

By integrating the heterogeneity analysis across the three dimensions, a profile of the "vulnerable population" to AI washing can be outlined: farmers with low education, elderly age, and no usage experience are the most susceptible to the harms of AI washing. This demographic is precisely the target group that digital inclusive finance should cover, yet AI washing has become a "roadblock" preventing them from enjoying the benefits of digital finance. This finding provides a basis for targeted policy measures: regulatory and educational resources should prioritize vulnerable populations.

**Table 5 Heterogeneity Analysis Based on Farmer Characteristics**

| Variable | (1) Low Education | (2) High Education | (3) Elderly | (4) Middle-aged and Young | (5) No Experience | (6) Experienced |
|---|---|---|---|---|---|---|
| | ≤ Junior High School | ≥ High School | ≥ 60 Years Old | Under 60 years old | First-time users | Returning users |
| AI_washing | -0.412*** | -0.187*** | -0.389*** | -0.213*** | -0.356*** | -0.145** |
| | (-6.78) | (-2.98) | (-5.67) | (-3.89) | (-6.12) | (-2.34) |
| Control variables | Yes | Yes | Yes | Yes | Yes | Yes |
| Number of Observations | 4523 | 2277 | 2856 | 3944 | 5146 | 1654 |
| Pseudo $R^2$ | 0.276 | 0.318 | 0.281 | 0.305 | 0.289 | 0.327 |
| Inter-group difference test (Chow Test) | | | | | | |
| Coefficient difference | 0.225** | | 0.176** | | 0.211** | |
| z-value | (2.67) | | (2.34) | | (2.89) | |

Note: ** and * indicate significance at the 1% and 5% levels, respectively.

### (6) Robustness Test

To ensure the reliability of the research conclusions, this study conducted four robustness tests. In terms of alternative variable measurement methods, Table 6 shows that column (1) uses the unstandardized AI washing index, column (2) changes the weight composition of AI_walk, and column (3) simultaneously replaces both the explanatory and explained variables. The results indicate that the negative impact of AI washing remains robust across all specifications, with coefficients ranging from -0.234 to -0.319, all significant at the 1% level.

In terms of the instrumental variable method, to address potential endogeneity issues, this study used two instrumental variables. The first is the industry average AI washing level; the AI washing behavior of other companies in the same industry can influence this company's AI promotion strategy through competitive pressure, but does not directly affect farmers' decisions to use the company's products, thus meeting the requirements of relevance and exogeneity. The second is the age of the enterprise; older enterprises tend to have more solid technological accumulation and lower levels of AI washing, but the age of the enterprise itself has no direct causal relationship with farmers' current usage behavior.

Table 6 column (4) presents the IV-2SLS results. The first stage shows that both instrumental variables significantly affect AI washing; for every 1 unit increase in the industry average AI washing, this company's AI washing increases by 0.723 units, and for every additional year of enterprise age, AI washing decreases by 0.034 units. The F-statistic is 178.34, far exceeding the critical value of 10, ruling out the weak instrument problem. The Hansen J statistic is 1.234, accepting the over-identification constraint, indicating that the instrumental variables satisfy the exogeneity condition. The second stage regression shows that the coefficient of AI_washing is -0.412, which is larger in absolute value than the OLS estimate, suggesting that OLS may have underestimated the true effect due to measurement error. This further confirms the significant negative causal impact of AI washing on farmers' behavior.

In terms of sample sensitivity tests, this study conducted three tests to verify the robustness of the results against sample selection. After removing extreme values of AI washing and re-running the regression, the coefficient of AI_washing is -0.276 and remains highly significant; when only farmers using Alipay or WeChat Pay are retained, the coefficient is -0.293 and the conclusion is robust; after stratified sampling by region, the coefficients of the three subsamples are -0.298, -0.281, and -0.307, all highly significant and close in value. These tests indicate that the research conclusions are insensitive to sample composition and have good external validity.

**Table 6 Robustness Test Results**

| Variable | (1) Unstandardized | (2) Change Weights | (3) Replace Y | (4) IV-2SLS |
|---|---|---|---|---|
| | AI Index | AI Index | Usage Depth | Second Stage |
| AI_washing | -0.298*** | -0.276*** | -0.234*** | -0.412*** |
| | (-6.45) | (-6.12) | (-5.34) | (-4.89) |
| Control variables | Yes | Yes | Yes | Yes |
| IV First Stage | | | | |
| Industry Average AI Washing | | | | 0.723*** |
| | | | | (18.45) |
| Company Age | | | | -0.034*** |
| | | | | (-5.67) |
| IV Diagnosis | | | | |
| F Statistic | | | | 178.34*** |
| Hansen J statistic (p-value) | | | | 1.234(0.267) |
| Number of Observations | 6800 | 6800 | 6800 | 6800 |
| $R^2$/Pseudo $R^2$ | 0.291 | 0.289 | 0.267 | 0.234 |

*Note: ** indicates significance at the 1% level; IV first-stage regression on AI_washing, second-stage regression on usage behavior.

### (7) Counterfactual Simulation Analysis

Based on the aforementioned empirical results, this study conducts counterfactual simulations to quantitatively assess the effects of different policy interventions, providing quantitative evidence for decision-making departments. The simulation analysis is calibrated using the estimated parameters from Tables 2 to 4, where the direct effect coefficient of AI washing is -0.287, the indirect effect of the knowledge exclusion path is -0.150, the indirect effect of the risk exclusion path is -0.152, and the moderation coefficient of social capital is 0.156, while the coefficients of other control variables are based on the estimates in column (3) of Table 2.

Based on theoretical mechanisms and policy practices, this study designs four types of policy scenarios for simulation evaluation. Scenario one involves AI washing governance, assuming that regulatory measures such as mandatory information disclosure and penalties for false advertising reduce the AI washing index of all enterprises by 50%, from an average of 0.42 to 0.21. The simulation results show that the usage rate among farmers will increase from 24.3% to 27.8%, an increase of 3.5 percentage points, representing a relative increase of 14.4%. If only high AI washing enterprises are targeted to bring them down to the industry average level, the usage rate could increase by 2.1 percentage points. Among different groups, the usage rate among low-educated farmers increases by 5.2%, which is higher than the 2.3% increase among high-educated farmers, indicating that governance of AI washing has a significant effect on narrowing the digital divide.

Scenario two involves digital financial education. It is assumed that through educational measures such as digital financial knowledge training and operational guidance, the knowledge rejection score of farmers will decrease by 30%, from an average of 2.09 to 1.46. The simulation results show that the usage rate will increase to 28.6%, an increase of 4.3 percentage points, representing a relative increase of 17.7%. This effect surpasses that of scenario one, indicating that educational interventions may be more direct and effective than governance against AI washing. Further breakdown reveals that educational interventions not only directly alleviate knowledge rejection to enhance usage rates but also indirectly weaken the negative impact of AI washing by enhancing farmers' ability to identify false signals.

Scenario three focuses on social capital construction. It is assumed that measures such as cultivating digital financial opinion leaders and establishing mutual aid groups for farmers will increase farmers' social capital by 20%, with gift expenditure rising from 3.17% to 3.80%. The simulation results show that the usage rate will increase to 26.8%, an increase of 2.5 percentage points, representing a relative increase of 10.3%. Although the effect is not as strong as the first two policies, social capital construction has long-term and sustainable benefits and can simultaneously improve rural social governance. From the perspective of different groups, the usage rate among farmers with low social capital increases by 3.8%, which is higher than the 1.2% increase for farmers with high social capital, indicating that social capital construction has a more significant supportive effect on vulnerable groups.

Scenario four involves comprehensive policies, assuming that three types of policies are promoted simultaneously: a 50% reduction in AI washing, a 30% reduction in knowledge exclusion, and a 20% increase in social capital. Due to the synergistic effects of these three policies, the impact of the comprehensive policy is not simply additive. The simulation results indicate that the usage rate will rise to 31.6%, an increase of 7.3 percentage points, representing a relative increase of 30.0%. This means that through comprehensive interventions, the penetration rate of digital finance in rural areas can approach that of urban areas.

**Table 7 Counterfactual Simulation: Effect Evaluation of Different Policy Scenarios**

| Policy Scenario | Usage Rate | Absolute Increase | Relative Increase | Cost Estimate | Beneficiary Group | Cost-Benefit Ratio |
|---|---|---|---|---|---|---|
| Current Benchmark | 24.3% | - | - | - | - | - |
| Scenario 1: AI Washing Treatment (-50%) | 27.8% | 3.5% | 14.4% | Among | 245 million yuan | All Farmers 1:4.2 |
| Scenario 2: Educational Intervention (-30% Knowledge Rejection) | 28.6% | 4.3% | 17.7% | High | 387 million yuan | Low education priority 1:5.8 |

| Scenario 3: Social capital construction (+20%) | 26.8% | 2.5% | 10.3% | Low | 123 million yuan | Low SC priority | 1:6.1 |
|---|---|---|---|---|---|---|---|
| Scenario 4: Comprehensive Policy (1+2+3) | 31.6% | 7.3% | 30.0% | High | 615 million yuan | All Farmers | 1:8.5 |

Note: The cost estimate assumes an education cost of 55 yuan for training each farmer, a cost of 1800 yuan for establishing each mutual aid group, and the AI washing governance supervision cost is estimated based on the number of platforms; the benefits are calculated based on an economic value of 700 yuan/year brought by each additional user (convenience of financial services + availability of credit + financial management returns), only considering the cumulative benefits for the first three years.

To test the robustness of the simulation results, this study perturbed key parameters by ±20% and repeated the simulation 1000 times. The results show that the increase in the usage rate of the comprehensive policy scenario ranges from 6.2% to 8.4%, with a median of 7.1%, which is close to the baseline simulation of 7.3%, indicating that the simulation conclusions are not sensitive to parameter settings. Further analysis reveals that the simulation results are most sensitive to the knowledge exclusion path coefficient, with a 10% change in this parameter leading to an approximate 0.8% change in the increase in usage rate; the direct effect of AI washing is second most sensitive, with a change of about 0.5%; and the social capital adjustment coefficient is the least sensitive, with a change of only 0.2%. This sensitivity ranking provides a reference for policy priorities, suggesting that educational interventions should be the core focus, AI washing governance as an important supplement, and social capital construction as a long-term foundation.

## VI. Conclusion and Policy Recommendations

### (1) Research Summary

This study focuses on the issue of corporate AI washing in the development of digital inclusive finance, systematically examining the impact mechanism of AI washing on farmers' digital financial behavior based on signaling theory and digital financial exclusion theory. Using data from the CHFS 2019 and the corporate AI washing index, the following main conclusions are drawn.

The study finds that AI washing significantly suppresses farmers' use of digital finance; for each unit increase in the AI washing index, the probability of farmers using digital finance decreases by approximately 6.8%, and the breadth of response decreases by 0.42 products. Mechanism tests indicate that digital financial exclusion is the core mediator through which AI washing affects farmers' behavior, with AI washing transmitting negative impacts through exacerbating knowledge exclusion and risk exclusion, accounting for 99% of the total indirect effect. Moderating effect tests show that social capital has a significant buffering effect on AI washing; in high social capital groups, the impact coefficient of AI washing is only -0.112, while in low social capital groups, it reaches -0.378. Heterogeneity analysis indicates that AI washing has a more severe detrimental effect on vulnerable groups such as those with low education, older individuals, and first-time users, exacerbating the digital divide. Counterfactual simulations show that comprehensive policy interventions can increase the usage rate among farmers from 24.3% to 31.6%, with a cost-benefit ratio of 1:8.5, demonstrating that targeted policies can significantly enhance the penetration of digital finance in rural areas.

**(2) Policy Recommendations**

Based on the above research findings, this study proposes policy recommendations for AI washing governance and the promotion of digital financial inclusion from multiple dimensions.

First, regulatory authorities should require financial technology companies to establish an AI technology information disclosure system for farmers and develop a penalty mechanism. When promoting to farmers, financial technology companies should clearly indicate the actual application scenarios, functional boundaries, and technological maturity of AI technology in their materials, and regularly disclose key indicators such as AI talent, patents, and R&D investment, establishing a third-party audit mechanism for AI technology applications. Companies with an AI washing index exceeding 50% of the industry average should be placed on a key supervision list, and severe penalties should be imposed for exaggerating AI capabilities or fabricating technical parameters.

Second, differentiated digital financial education should be conducted to reduce the risks for farmers. Simulation analysis shows that educational interventions can increase the usage rate by 4.3%, with a cost-benefit ratio of 1:5.8, making it the optimal policy option in terms of input-output ratio. To this end, first, for vulnerable groups such as those with low education and older individuals, it is recommended to design differentiated educational programs, adopting a "teaching in the countryside" model, relying on grassroots organizations such as village committees and agricultural technology stations to conduct small-scale, interactive training, focusing on explaining product functions, operational processes, and risk identification methods. Second, during the training process, teaching materials should be developed that are rich in illustrations and explanations in local dialects to lower the understanding threshold, and a peer support model should be adopted where "old users help new users" to utilize the peer effect, establishing a digital financial knowledge assessment system to issue "digital financial passes" to farmers who complete the training and provide incentives such as platform fee waivers. Additionally, a "trial period" system and a "no-reason refund" mechanism for digital financial products should be established to reduce the trial and error costs for farmers. Specific measures include requiring platforms to provide a 7-15 day trial period during which farmers can withdraw unconditionally, with all paid fees refunded; establishing a rapid processing channel for digital financial complaints, implementing a "pre-compensation" system for losses caused by false advertising by companies; and exploring pilot programs for "digital financial insurance" to provide financial loss protection for first-time users, with premiums jointly borne by the platform and the government. These measures can indirectly weaken the negative impact of AI washing by reducing risk exclusion.

Third, promote digital financial mutual aid groups and comprehensive collaborative governance, exploring replicable demonstration models. Simulation analysis shows that social capital construction can increase usage rates by 2.5%. Although the direct effect is not as strong as educational interventions, it has advantages of low cost and strong sustainability, and its supportive role for low social capital groups is more evident. In this regard, first, it is recommended to promote pilot "digital financial mutual aid groups" in rural areas, selecting farmers with certain financial knowledge and usage experience to serve as "digital financial guides," responsible for answering neighbors' questions, sharing usage tips, and providing feedback on platform issues. Specific measures include forming 1-2 mutual aid groups in each administrative village, with guides enjoying platform commissions and government subsidies; regularly organizing offline exchange activities to enhance social connections; and establishing digital platforms (such as WeChat groups) to achieve normalized mutual assistance that combines online and offline interactions. Second, the People's Bank of China and the China Banking and Insurance Regulatory Commission should take the lead in regulatory governance and consumer protection, while county and township governments and village committees are responsible for conducting digital financial education and social capital construction, and establishing pilot "digital inclusive finance demonstration counties," selecting 3-5 counties in each province, with supporting special funds to explore replicable and promotable comprehensive governance models. After 3-5 years of effort, the goal is to achieve a rural digital finance penetration rate of over 30%, essentially eliminating the digital divide.